# Forecasting the Price of Rice in Banda Aceh after Covid-19

Fadhlul Mubarak[1], Vinny Yuliani Sundara[2], Nurniswah[3]

*[1]Assistant Professor,*
*Department of Agribusiness,*
*Faculty of Agriculture, Jambi University, 36361, Jambi, Indonesia.*
*[2]Assistant Professor,*
*Department of Statistics,*
*Faculty of Science and Technology, UIN Sulthan Thaha Saifuddin Jambi, 36361, Jambi, Indonesia.*
*[3]Assistant Professor,*
*Department of Social Science Education,*
*Faculty of Education, UIN Fatmawati Sukarno Bengkulu, 38211, Bengkulu, Indonesia.*
*Corresponding Author: Fadhlul Mubarak.*

## ABSTRACT

*This research aims to predict the price of rice in Banda Aceh after the occurrence of Covid-19. The last observation carried forward (LOCF) imputation technique has been used to solve the problem of missing values from this research data. Furthermore, the technique used to forecast rice prices in Banda Aceh is auto-ARIMA which is the best ARIMA model based on AIC, AICC, or BIC values. The results of this research show that the ARIMA model (0,0,5) is the best model to predict the prices of lower quality rice I (BKB1), lower quality rice II (BKB2), medium quality rice I (BKM1), medium quality rice II (BKM2), super quality rice I (BKS1), and super quality rice II (BKS2). Based on this model, the results of forecasting rice prices for all qualities show that there was a decline for some time (between September 1, 2023 and September 6, 2023) and then remained constant (between September 6, 2023 and December 31, 2023).*

---

---

## I.    INTRODUCTION

Banda Aceh is the capital of Aceh Province which is located in the western part of Indonesia. The city is one of the areas that enforces Islamic Sharia in Indonesia [1]–[4]. Of course, this makes this city unique for conducting research. This research aims to look at the price of rice in the city after the outbreak of Covid-19. Rice is the staple food of the majority of the Indonesian population, especially Banda Aceh. However, there is still very little research that discusses the relationship between rice prices and Banda Aceh, especially forecasting rice prices. Putri et al. (2021) measured the level of consumer satisfaction with the service and prices of products assisted by the Basic Food Program in Banda Aceh, where one of the assistances is rice. As a result, respondents felt satisfied because the service was quite good and reduced the burden on respondents' expenses. Furthermore, Nelly et al. on 2018 analyzed the factors that influence rice price fluctuations in Aceh Province. The research looks at the simultaneous relationship between price, production and consumption. The research results show that the price is influenced by the price of grain, rice production, and the highest retail price (HET). Apart from that, Pradana (2019) analyzed changes and volatility in prices of strategic food commodities, one of which is rice, and their impact on inflation in Banda Aceh. The result is price volatility in changes in rice prices. Then, the regression estimation results show that changes in rice prices have a significant positive effect on inflation in Banda Aceh. However, rice price volatility does not have a significant effect on inflation in Banda Aceh.

Various statistical models can be used for forecasting, including ARIMA [8]–[13], neural network [14]–[18], moving average smoothing [19], etc. Specifically, Ohyver & Pudjihastuti on 2018 predicted the price of medium quality rice to anticipate price fluctuations using the autoregressive integrated moving average (ARIMA) model. This research is the basis of this research but still has several differences. The first difference is the location where the research used 34 provinces in Indonesia while this research used Banda Aceh as the research location. Second is the data source, this research uses data sources from the Ministry of Trade while this research uses sources from Bank Indonesia. The third is the imputation technique, this study does not explain cases of missing values in the data, whereas in this study there were cases of missing values so imputation was necessary to resolve them. The missing value technique has been widely applied to various research objects, including genotypes [20], management of water-quality data [21], production data of dairy cattle [22], US crossbred dairy cattle [23], low-coverage sequencing in Holstein cattle [24], etc. Fourth, the model used in this research is ARIMA in general, whereas in this research the auto-ARIMA technique is used. And finally, this study used the price of medium quality rice, while this study used lower quality rice I (BKB1), lower quality rice II (BKB2), medium quality rice I (BKM1), medium quality rice II (BKM2), super quality rice I (BKS1), and super quality rice II (BKS2).

In the method section, the data source, location and time used in this research will be explained. Apart from that, there is an explanation of the imputation process to overcome missing values contained in the data and ARIMA models. In the results and analysis section there is data on rice prices and the results of the imputation. Next, the best models for forecasting rice prices for each rice quality are displayed. Finally, the conclusion section of this research and suggestions for future research.

## II. METHOD

This research uses rice price data at the Banda Aceh Traditional Market from the National Strategic Food Price Information Center (PIHPS Nasional) between December 31, 2019 and August 31, 2023. The time interval is based on the time when Covid-19 was declared until the time this research began. This data is daily data on rice prices in Banda Aceh traditional markets. Also, this research used lower quality rice I (BKB1), lower quality rice II (BKB2), medium quality rice I (BKM1), medium quality rice II (BKM2), super quality rice I (BKS1), and super second quality rice. (BKS2) as the variable studied. However, these data have missing values at some times. So, this research solves this problem by using the last observation missing value (LOCF) imputation method which follows **equation 1**.

$$A_i^* = \begin{cases} if\ complete\ data\ then\ A_i^* = A_i \\ if\ missing\ data\ then\ \ A_i^* = A_{i-e} \end{cases} \tag{1}$$

where $A_i^*$ is the result of imputation of rice prices and $A_{i-a}$ is a recursive process of finding data lags until no missing values occur. After the missing values in the rice price data have been resolved, they are modeled using the ARIMA model which follows **equation 2** and **equation 3**.

$$\nabla A_i = c + \sum_{e=1}^{e=p} \phi_e \nabla A_{i-e} + \varepsilon_i - \sum_{f=1}^{f=q} \theta_f \varepsilon_{i-f} \tag{2}$$

$$\left(1 - \phi_1 B - \cdots - \phi_p B^p\right)(1-B)^d A_i = c + \left(1 + \theta_1 B + \cdots + \theta_q B^q\right)\varepsilon_i \tag{3}$$

where $\nabla A_i = (1-B)^d A_i$ is dependent variable, $c$ is konstanta, $\phi_e$ is parameter for autoregressive, $\nabla A_{i-e}$ is lag of dependent variable, $\theta_f$ is parameter for moving average, $\varepsilon_i$ is error, and $\varepsilon_{i-f}$ is lag of error. Next, $p$ is the number of lag dependent variable, $d$ is the degree of differencing, and $q$ is the order of the moving average. Furthermore, the ARIMA model developed into an auto-ARIMA technique which is the best ARIMA model based on either AIC (**equation 4**), AICC (**equation 5**) or BIC value (**equation 6**).

$$AIC(A_i) = 2k - 2\ ln(\hat{L}) \tag{4}$$

where $k$ be the number of estimated parameters in the model and $\hat{L}$ be the maximized value of the likelihood function for the model.

$$AICC(A_i) = AIC(A_i) + \frac{2k^2 + 2k}{n - k - 1} \tag{5}$$

where $n$ denotes the sample size and $k$ denotes the number of parameters

$$BIC(A_i) = k\ ln(n) - 2\ ln(\hat{L}) \tag{6}$$

## III. RESULTS AND ANALYSIS

The data used in this research consists of 7 columns and 958 rows. The first column is daily time information, the second column is lower quality rice I (BKB1), the third column is lower quality rice II (BKB2), the fourth column is medium quality rice I (BKM1), the fifth column is medium quality rice II (BKM2), the sixth column is super quality rice I (BKS1), and the seventh column is super quality rice II (BKS2). The head and tail of the rice price data in Banda Aceh are available in **Table 1** and **Table 2**. Based on **Table 1**, there are several daily rice price data in Banda Aceh that are not recorded, including January 1, 2020, January 4, 2020, and January 5, 2020. So Also in **Table 2**, there is some daily data on rice prices in Banda Aceh that is not recorded, including August 26, 2023 and August 27, 2023.

**Table 1**. Head of rice price data in Banda Aceh

| Date | BKB1 | BKB2 | BKM1 | BKM2 | BKS1 | BKS2 |
|---|---|---|---|---|---|---|
| 2019-12-31 | 9850 | 10000 | 10250 | 10300 | 11900 | 11150 |
| 2020-01-01 | NA | NA | NA | NA | NA | NA |
| 2020-01-02 | 9850 | 10000 | 10250 | 10300 | 11900 | 11150 |
| 2020-01-03 | 9850 | 10000 | 10250 | 10300 | 11900 | 11150 |
| 2020-01-06 | 10000 | 10150 | 10400 | 10400 | 12150 | 11400 |
| 2020-01-07 | 10000 | 10150 | 10400 | 10400 | 12150 | 11400 |

**Table 2**. Tail of rice price data in Banda Aceh

| Date | BKB1 | BKB2 | BKM1 | BKM2 | BKS1 | BKS2 |
|---|---|---|---|---|---|---|
| 2023-08-24 | 11600 | 12200 | 12100 | 12100 | 13000 | 12900 |
| 2023-08-25 | 11600 | 12200 | 12100 | 12100 | 13000 | 12900 |
| 2023-08-28 | 11600 | 12200 | 12150 | 12100 | 13000 | 12900 |
| 2023-08-29 | 11600 | 12200 | 12150 | 12100 | 13900 | 12950 |
| 2023-08-30 | 11600 | 12200 | 12150 | 12100 | 13900 | 12950 |
| 2023-08-31 | 11600 | 12200 | 12150 | 12100 | 13900 | 12950 |

In connection with this problem, LOCF imputation is applied to rice price data in Banda Aceh. So, the complete data is available in Table 3 and Table 4. In data that has not been imputed, the minimum value of the price of lower quality rice I is 9050 IDR, the price of lower quality rice II is 9500 IDR, the price of medium quality I rice is 9800 IDR, the price medium quality rice II is 9800 IDR, super quality rice I is 11350 IDR, and super quality rice II is 10650 IDR. Furthermore, the maximum value of the price of lower quality rice I is 11600 IDR, the price of lower quality rice II is 12200 IDR, the price of medium quality rice I is 12150 IDR, the price of medium quality rice II is 12100 IDR, the price of super quality rice I is 13900 IDR and the price of super II quality rice is 12950 IDR.

**Table 3**. Head of rice price data in Banda Aceh (after imputation)

| Date | BKB1 | BKB2 | BKM1 | BKM2 | BKS1 | BKS2 |
|---|---|---|---|---|---|---|
| 2019-12-31 | 9850 | 10000 | 10250 | 10300 | 11900 | 11150 |
| 2020-01-01 | 9850 | 10000 | 10250 | 10300 | 11900 | 11150 |
| 2020-01-02 | 9850 | 10000 | 10250 | 10300 | 11900 | 11150 |
| 2020-01-03 | 9850 | 10000 | 10250 | 10300 | 11900 | 11150 |
| 2020-01-04 | 9850 | 10000 | 10250 | 10300 | 11900 | 11150 |
| 2020-01-05 | 9850 | 10000 | 10250 | 10300 | 11900 | 11150 |

**Table 4**. Tail of rice price data in Banda Aceh (after imputation)

| Date | BKB1 | BKB2 | BKM1 | BKM2 | BKS1 | BKS2 |
|---|---|---|---|---|---|---|
| 2023-08-26 | 11600 | 12200 | 12100 | 12100 | 13000 | 12900 |
| 2023-08-27 | 11600 | 12200 | 12100 | 12100 | 13000 | 12900 |
| 2023-08-28 | 11600 | 12200 | 12150 | 12100 | 13000 | 12900 |
| 2023-08-29 | 11600 | 12200 | 12150 | 12100 | 13900 | 12950 |
| 2023-08-30 | 11600 | 12200 | 12150 | 12100 | 13900 | 12950 |
| 2023-08-31 | 11600 | 12200 | 12150 | 12100 | 13900 | 12950 |

Meanwhile, after the rice price data was imputed, the price data for lower quality I rice on January 1, 2020 was 9850 IDR, lower second quality rice was 10000 IDR, medium I quality rice was 10250 IDR, medium 2 quality rice was 10300 IDR, super I quality rice is 11900 IDR and super quality II rice is 11150 IDR. With the LOCF imputation technique, the price of rice on January 1 2020 is the same as on December 31, 2019. Likewise, the price of rice on January 4, 2020 is the same as on January 3, 2020, January 5, 2020 is the same as on January 4, 2020. This is the same in the tail part of data, on August 26, 2023 the same as on August 25, 2023 and on August 27, 2023 the same as on August 26, 2023.

$$BKB1_i = 9995.01 + \varepsilon_i + 2.08\varepsilon_{i-1} + 2.74\varepsilon_{i-2} + 2.52\varepsilon_{i-3} + 1.56\varepsilon_{i-4} + 0.58\varepsilon_{i-5} \qquad (7)$$

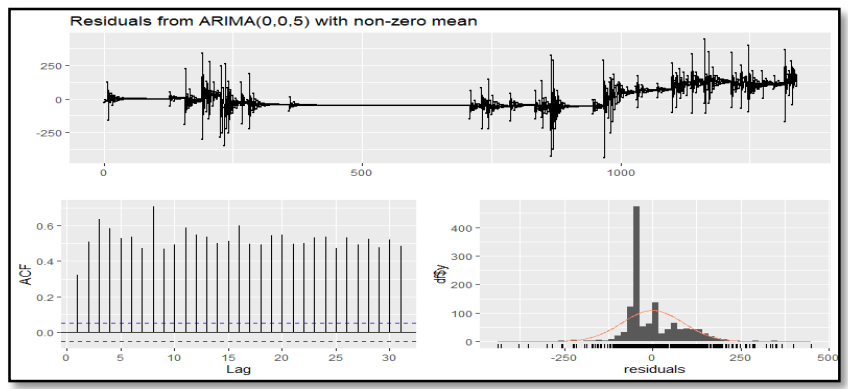

**Figure 1**. Residuals testing from BKB1 variable

$$BKB2_i = 10422.90 + \varepsilon_i + 1.73\varepsilon_{i-1} + 2.11\varepsilon_{i-2} + 1.97\varepsilon_{i-3} + 1.37\varepsilon_{i-4} + 0.62\varepsilon_{i-5} \qquad (8)$$

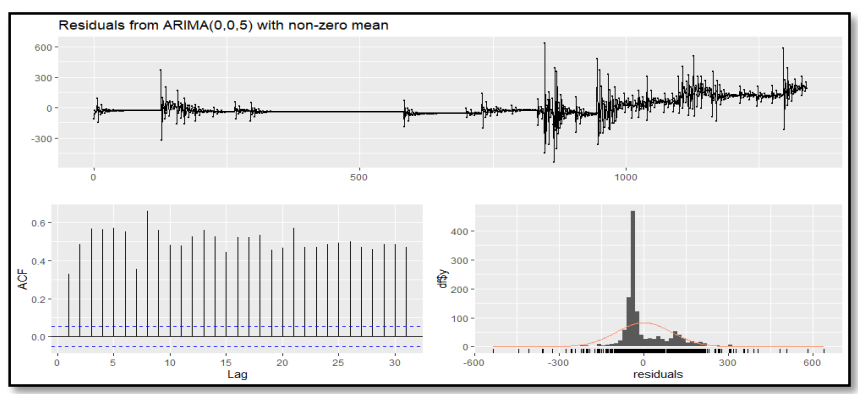

**Figure 2**. Residuals testing from BKB2 variable

$$BKM1_i = 10642.36 + \varepsilon_i + 1.84\varepsilon_{i-1} + 2.39\varepsilon_{i-2} + 2.13\varepsilon_{i-3} + 1.34\varepsilon_{i-4} + 0.52\varepsilon_{i-5} \qquad (9)$$

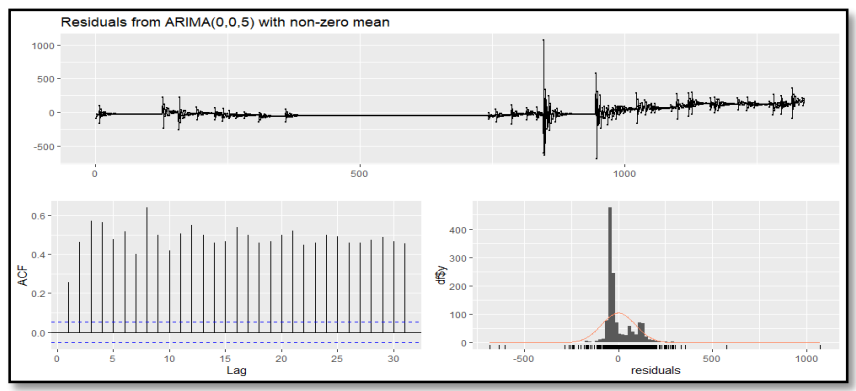

**Figure 3**. Residuals testing from BKM1 variable

$$BKM2_i = 10678.66 + \varepsilon_i + 1.83\varepsilon_{i-1} + 2.33\varepsilon_{i-2} + 2.10\varepsilon_{i-3} + 1.35\varepsilon_{i-4} + 0.52\varepsilon_{i-5} \qquad (10)$$

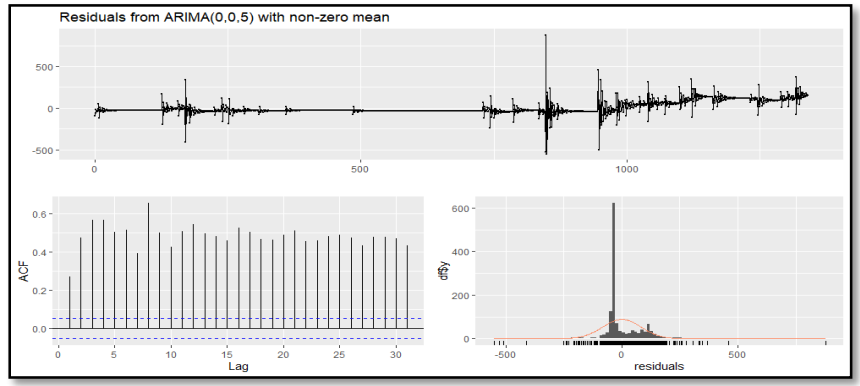

**Figure 4**. Residuals testing from BKM2 variable

$$BKS1_i = 12313.12 + \varepsilon_i + 1.63\varepsilon_{i-1} + 1.95\varepsilon_{i-2} + 1.79\varepsilon_{i-3} + 1.29\varepsilon_{i-4} + 0.60\varepsilon_{i-5} \tag{11}$$

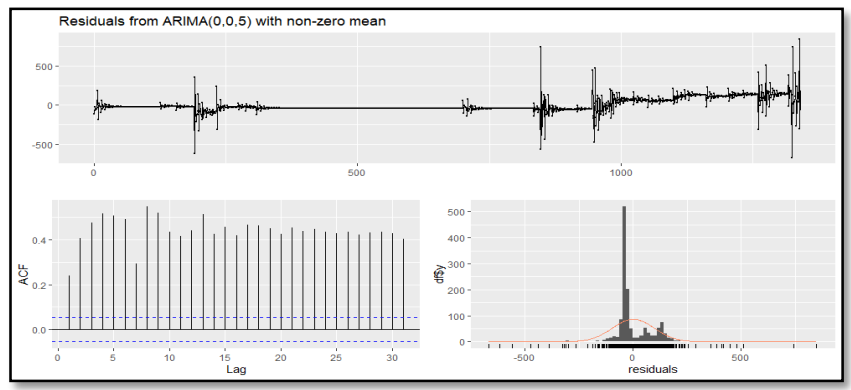

**Figure 5**. Residuals testing from BKS1 variable

$$BKS2_i = 11387.50 + \varepsilon_i + 2.08\varepsilon_{i-1} + 2.79\varepsilon_{i-2} + 2.54\varepsilon_{i-3} + 1.59\varepsilon_{i-4} + 0.60\varepsilon_{i-5} \tag{12}$$

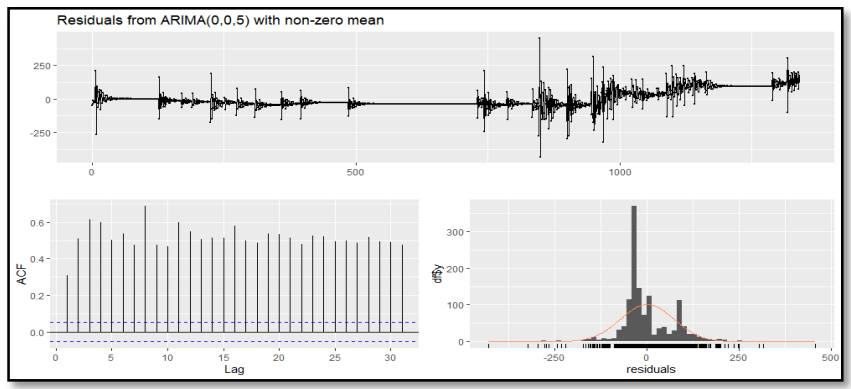

**Figure 6**. Residuals testing from BKS1 variable

Based on the ARIMA model technique (0,0,5) is the best model for forecasting rice prices for all qualities, including lower quality rice I, lower quality rice II, medium quality rice I, medium quality rice II, super quality rice I, and super quality rice II rice. super II. **Equation 7** represented the best forecasting model for lower quality rice I, **equation 8** represented the best forecasting model for lower quality rice II, **equation 9** represented the best forecasting model for medium quality rice I, **equation 10** represented the best forecasting model for medium quality rice II, **equation 11** represented the best forecasting model for super quality rice I and **equation 12** represented the best forecasting model for super quality rice II. Apart from that, residual testing for each model is available in **Figure 1-6**. Based on the residuals testing, the model is suitable for use in forecasting rice prices for all qualities in Banda Aceh.

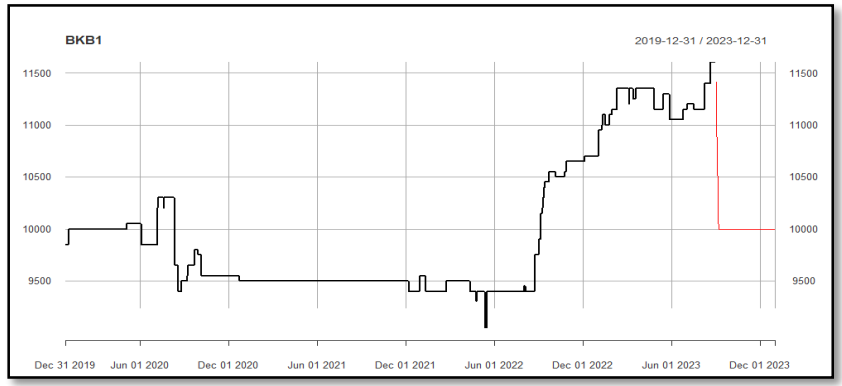

**Figure 7**. Forecasting the price of lower quality rice I

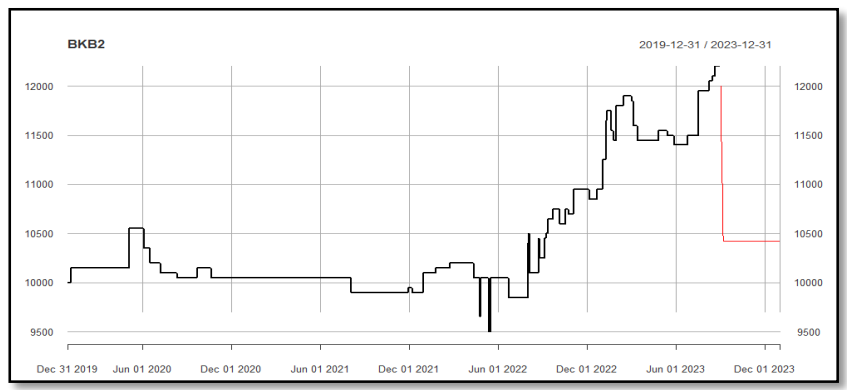

**Figure 8**. Forecasting the price of lower quality rice II

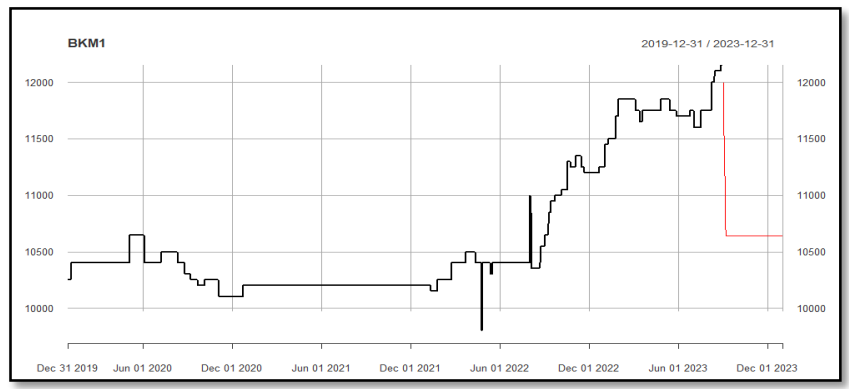

**Figure 9**. Forecasting the price of medium quality rice I

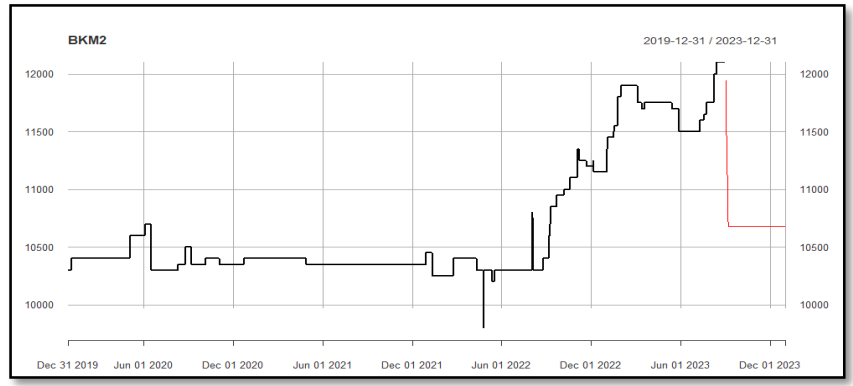

**Figure 10**. Forecasting the price of medium quality rice II

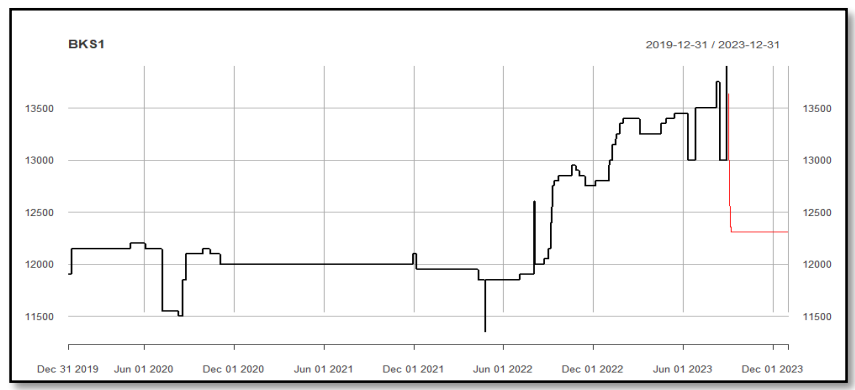

**Figure 11**. Forecasting the price of super quality rice I

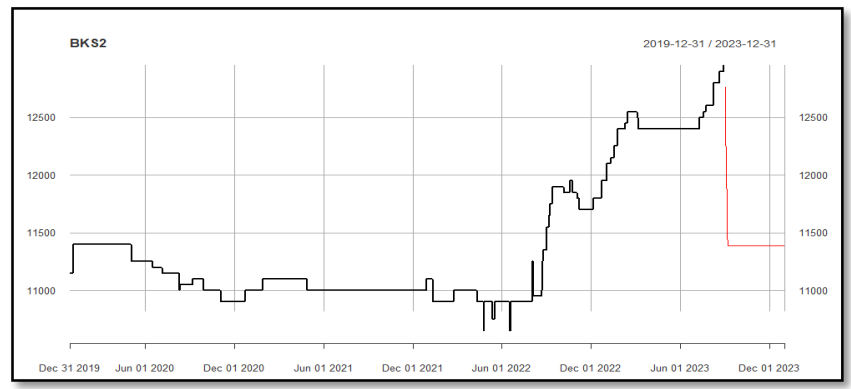

**Figure 12**. Forecasting the price of super quality rice II

Based on the results of forecasting rice prices for all qualities in **Figure 7-12**, it can be seen that there was a decline for some time and then remained constant, in general. The results of forecasting the price of lower quality rice I are 11410 IDR, 11081 IDR, 10697 IDR, 10328 IDR, 10099 IDR, and 9995 IDR respectively between September 1, 2023 and September 6, 2023. Furthermore, the results of forecasting the price of lower quality rice I between September 7, 2023 and December 31, 2023 constant at 9995 IDR. In addition, the forecast results for the price of lower quality rice II on September 1, 2023 is 12005 IDR, on September 2, 2023 it is 11655 IDR, on September 3, 11221 IDR, on September 4, 2023 it is 10803 IDR, on September 5, 2023 it is 10539 IDR, on September 6, 2023 is 10422 IDR. After that, the forecast results for the price of lower quality rice I between September 7, 2023 and December 31, 2023 are the same as the price on September 6, 2023.

**Table 5**. Results of rice price forecasting in September 2023

| Date | BKB1 | BKB2 | BKM1 | BKM2 | BKS1 | BKS2 |
|------|------|------|------|------|------|------|
| 2023-09-01 | 11411 | 12005 | 11995 | 11945 | 13640 | 12761 |
| 2023-09-02 | 11081 | 11655 | 11720 | 11652 | 13244 | 12441 |
| 2023-09-03 | 10698 | 11221 | 11343 | 11287 | 12757 | 12060 |
| 2023-09-04 | 10329 | 10804 | 10998 | 10966 | 12359 | 11705 |
| 2023-09-05 | 10099 | 10539 | 10747 | 10758 | 12420 | 11491 |
| 2023-09-06 | 9995 | 10423 | 10642 | 10679 | 12313 | 11388 |
| 2023-09-07 | 9995 | 10423 | 10642 | 10679 | 12313 | 11388 |
| 2023-09-08 | 9995 | 10423 | 10642 | 10679 | 12313 | 11388 |
| 2023-09-09 | 9995 | 10423 | 10642 | 10679 | 12313 | 11388 |
| 2023-09-10 | 9995 | 10423 | 10642 | 10679 | 12313 | 11388 |
| 2023-09-11 | 9995 | 10423 | 10642 | 10679 | 12313 | 11388 |
| 2023-09-12 | 9995 | 10423 | 10642 | 10679 | 12313 | 11388 |
| 2023-09-13 | 9995 | 10423 | 10642 | 10679 | 12313 | 11388 |
| 2023-09-14 | 9995 | 10423 | 10642 | 10679 | 12313 | 11388 |
| 2023-09-15 | 9995 | 10423 | 10642 | 10679 | 12313 | 11388 |
| 2023-09-16 | 9995 | 10423 | 10642 | 10679 | 12313 | 11388 |
| 2023-09-17 | 9995 | 10423 | 10642 | 10679 | 12313 | 11388 |
| 2023-09-18 | 9995 | 10423 | 10642 | 10679 | 12313 | 11388 |
| 2023-09-19 | 9995 | 10423 | 10642 | 10679 | 12313 | 11388 |
| 2023-09-20 | 9995 | 10423 | 10642 | 10679 | 12313 | 11388 |
| 2023-09-21 | 9995 | 10423 | 10642 | 10679 | 12313 | 11388 |
| 2023-09-22 | 9995 | 10423 | 10642 | 10679 | 12313 | 11388 |
| 2023-09-23 | 9995 | 10423 | 10642 | 10679 | 12313 | 11388 |
| 2023-09-24 | 9995 | 10423 | 10642 | 10679 | 12313 | 11388 |
| 2023-09-25 | 9995 | 10423 | 10642 | 10679 | 12313 | 11388 |
| 2023-09-26 | 9995 | 10423 | 10642 | 10679 | 12313 | 11388 |
| 2023-09-27 | 9995 | 10423 | 10642 | 10679 | 12313 | 11388 |
| 2023-09-28 | 9995 | 10423 | 10642 | 10679 | 12313 | 11388 |
| 2023-09-29 | 9995 | 10423 | 10642 | 10679 | 12313 | 11388 |
| 2023-09-30 | 9995 | 10423 | 10642 | 10679 | 12313 | 11388 |

**Tabel 6**. Results of rice price forecasting in October 2023

| Date | BKB1 | BKB2 | BKM1 | BKM2 | BKS1 | BKS2 |
|------|------|------|------|------|------|------|
| 2023-10-01 | 9995 | 10423 | 10642 | 10679 | 12313 | 11388 |
| 2023-10-02 | 9995 | 10423 | 10642 | 10679 | 12313 | 11388 |
| 2023-10-03 | 9995 | 10423 | 10642 | 10679 | 12313 | 11388 |
| 2023-10-04 | 9995 | 10423 | 10642 | 10679 | 12313 | 11388 |
| 2023-10-05 | 9995 | 10423 | 10642 | 10679 | 12313 | 11388 |
| 2023-10-06 | 9995 | 10423 | 10642 | 10679 | 12313 | 11388 |
| 2023-10-07 | 9995 | 10423 | 10642 | 10679 | 12313 | 11388 |
| 2023-10-08 | 9995 | 10423 | 10642 | 10679 | 12313 | 11388 |
| 2023-10-09 | 9995 | 10423 | 10642 | 10679 | 12313 | 11388 |
| 2023-10-10 | 9995 | 10423 | 10642 | 10679 | 12313 | 11388 |
| 2023-10-11 | 9995 | 10423 | 10642 | 10679 | 12313 | 11388 |
| 2023-10-12 | 9995 | 10423 | 10642 | 10679 | 12313 | 11388 |
| 2023-10-13 | 9995 | 10423 | 10642 | 10679 | 12313 | 11388 |
| 2023-10-14 | 9995 | 10423 | 10642 | 10679 | 12313 | 11388 |
| 2023-10-15 | 9995 | 10423 | 10642 | 10679 | 12313 | 11388 |
| 2023-10-16 | 9995 | 10423 | 10642 | 10679 | 12313 | 11388 |
| 2023-10-17 | 9995 | 10423 | 10642 | 10679 | 12313 | 11388 |
| 2023-10-18 | 9995 | 10423 | 10642 | 10679 | 12313 | 11388 |
| 2023-10-19 | 9995 | 10423 | 10642 | 10679 | 12313 | 11388 |
| 2023-10-20 | 9995 | 10423 | 10642 | 10679 | 12313 | 11388 |
| 2023-10-21 | 9995 | 10423 | 10642 | 10679 | 12313 | 11388 |
| 2023-10-22 | 9995 | 10423 | 10642 | 10679 | 12313 | 11388 |
| 2023-10-23 | 9995 | 10423 | 10642 | 10679 | 12313 | 11388 |
| 2023-10-24 | 9995 | 10423 | 10642 | 10679 | 12313 | 11388 |
| 2023-10-25 | 9995 | 10423 | 10642 | 10679 | 12313 | 11388 |
| 2023-10-26 | 9995 | 10423 | 10642 | 10679 | 12313 | 11388 |
| 2023-10-27 | 9995 | 10423 | 10642 | 10679 | 12313 | 11388 |
| 2023-10-28 | 9995 | 10423 | 10642 | 10679 | 12313 | 11388 |
| 2023-10-29 | 9995 | 10423 | 10642 | 10679 | 12313 | 11388 |
| 2023-10-30 | 9995 | 10423 | 10642 | 10679 | 12313 | 11388 |
| 2023-10-31 | 9995 | 10423 | 10642 | 10679 | 12313 | 11388 |

The price forecasting results for medium and super quality rice have the same pattern as lower quality rice. The price of medium I quality rice began to decline between September 1, 2023 and September 6, 2023 with values of 11994 IDR, 11719 IDR, 11343 IDR, 10997 IDR, 10747 IDR and 10643 IDR, respectively. Furthermore, the forecast results for the price of medium quality rice II in the same period are 11944 IDR, 11651 IDR, 11286 IDR, 10966 IDR, 10758 IDR, and 10678 IDR, respectively. Likewise with the forecasting results for the price of super quality rice I, where on September 1, 2023 it is 13639 IDR, September 2, 2023 is 13244 IDR, September 3, 2023 is 12757 IDR, September 4, 2023 is 12358 IDR, September 5, 2023 is 12419 IDR, and September 6, 2023 is 12313 IDR. The same thing, the results of forecasting the price of super quality rice II are 12761 IDR, 12441 IDR, 12060 IDR, 11704 IDR, 11704 IDR, 11490 IDR, 11387 IDR and 11387 IDR, in the same period respectively. Finally, complete forecasting results for rice prices for the period between September 1, 2023 and December 31, 2023 are available in **Table 5** - **Table 8**.

**Table 7**. Results of rice price forecasting in November 2023

| Date | BKB1 | BKB2 | BKM1 | BKM2 | BKS1 | BKS2 |
|------|------|------|------|------|------|------|
| 2023-11-01 | 9995 | 10423 | 10642 | 10679 | 12313 | 11388 |
| 2023-11-02 | 9995 | 10423 | 10642 | 10679 | 12313 | 11388 |
| 2023-11-03 | 9995 | 10423 | 10642 | 10679 | 12313 | 11388 |
| 2023-11-04 | 9995 | 10423 | 10642 | 10679 | 12313 | 11388 |
| 2023-11-05 | 9995 | 10423 | 10642 | 10679 | 12313 | 11388 |
| 2023-11-06 | 9995 | 10423 | 10642 | 10679 | 12313 | 11388 |
| 2023-11-07 | 9995 | 10423 | 10642 | 10679 | 12313 | 11388 |
| 2023-11-08 | 9995 | 10423 | 10642 | 10679 | 12313 | 11388 |
| 2023-11-09 | 9995 | 10423 | 10642 | 10679 | 12313 | 11388 |
| 2023-11-10 | 9995 | 10423 | 10642 | 10679 | 12313 | 11388 |
| 2023-11-11 | 9995 | 10423 | 10642 | 10679 | 12313 | 11388 |
| 2023-11-12 | 9995 | 10423 | 10642 | 10679 | 12313 | 11388 |
| 2023-11-13 | 9995 | 10423 | 10642 | 10679 | 12313 | 11388 |
| 2023-11-14 | 9995 | 10423 | 10642 | 10679 | 12313 | 11388 |
| 2023-11-15 | 9995 | 10423 | 10642 | 10679 | 12313 | 11388 |
| 2023-11-16 | 9995 | 10423 | 10642 | 10679 | 12313 | 11388 |
| 2023-11-17 | 9995 | 10423 | 10642 | 10679 | 12313 | 11388 |
| 2023-11-18 | 9995 | 10423 | 10642 | 10679 | 12313 | 11388 |
| 2023-11-19 | 9995 | 10423 | 10642 | 10679 | 12313 | 11388 |
| 2023-11-20 | 9995 | 10423 | 10642 | 10679 | 12313 | 11388 |
| 2023-11-21 | 9995 | 10423 | 10642 | 10679 | 12313 | 11388 |
| 2023-11-22 | 9995 | 10423 | 10642 | 10679 | 12313 | 11388 |
| 2023-11-23 | 9995 | 10423 | 10642 | 10679 | 12313 | 11388 |
| 2023-11-24 | 9995 | 10423 | 10642 | 10679 | 12313 | 11388 |
| 2023-11-25 | 9995 | 10423 | 10642 | 10679 | 12313 | 11388 |
| 2023-11-26 | 9995 | 10423 | 10642 | 10679 | 12313 | 11388 |
| 2023-11-27 | 9995 | 10423 | 10642 | 10679 | 12313 | 11388 |
| 2023-11-28 | 9995 | 10423 | 10642 | 10679 | 12313 | 11388 |
| 2023-11-29 | 9995 | 10423 | 10642 | 10679 | 12313 | 11388 |
| 2023-11-30 | 9995 | 10423 | 10642 | 10679 | 12313 | 11388 |

In summary, the forecast results for lower quality rice prices I for this time period are between 9995 IDR and 11411 IDR (**Figure 7**). Furthermore, the forecast results for lower quality rice prices II are between 10,423 IDR and 12,005 IDR (**Figure 8**). Apart from that, the forecast results for the price of medium quality rice I for this time period are between 10642 IDR and 11995 IDR (**Figure 9**). Also, the forecast results for the price of medium quality rice II in this period are between 10679 IDR and 11945 IDR (**Figure 10**). Furthermore, the forecast results for the price of super quality rice I in that period are between 12313 IDR and 13640 IDR (**Figure 11**). Finally, the forecast results for the price of super quality rice I at that time were between 11388 IDR and 12761 IDR (**Figure 12**).

**Tabel 8**. Results of rice price forecasting in December 2023

| Date | BKB1 | BKB2 | BKM1 | BKM2 | BKS1 | BKS2 |
|------|------|------|------|------|------|------|
| 2023-12-01 | 9995 | 10423 | 10642 | 10679 | 12313 | 11388 |
| 2023-12-02 | 9995 | 10423 | 10642 | 10679 | 12313 | 11388 |
| 2023-12-03 | 9995 | 10423 | 10642 | 10679 | 12313 | 11388 |
| 2023-12-04 | 9995 | 10423 | 10642 | 10679 | 12313 | 11388 |
| 2023-12-05 | 9995 | 10423 | 10642 | 10679 | 12313 | 11388 |
| 2023-12-06 | 9995 | 10423 | 10642 | 10679 | 12313 | 11388 |
| 2023-12-07 | 9995 | 10423 | 10642 | 10679 | 12313 | 11388 |
| 2023-12-08 | 9995 | 10423 | 10642 | 10679 | 12313 | 11388 |
| 2023-12-09 | 9995 | 10423 | 10642 | 10679 | 12313 | 11388 |
| 2023-12-10 | 9995 | 10423 | 10642 | 10679 | 12313 | 11388 |
| 2023-12-11 | 9995 | 10423 | 10642 | 10679 | 12313 | 11388 |
| 2023-12-12 | 9995 | 10423 | 10642 | 10679 | 12313 | 11388 |
| 2023-12-13 | 9995 | 10423 | 10642 | 10679 | 12313 | 11388 |
| 2023-12-14 | 9995 | 10423 | 10642 | 10679 | 12313 | 11388 |
| 2023-12-15 | 9995 | 10423 | 10642 | 10679 | 12313 | 11388 |
| 2023-12-16 | 9995 | 10423 | 10642 | 10679 | 12313 | 11388 |
| 2023-12-17 | 9995 | 10423 | 10642 | 10679 | 12313 | 11388 |
| 2023-12-18 | 9995 | 10423 | 10642 | 10679 | 12313 | 11388 |
| 2023-12-19 | 9995 | 10423 | 10642 | 10679 | 12313 | 11388 |
| 2023-12-20 | 9995 | 10423 | 10642 | 10679 | 12313 | 11388 |
| 2023-12-21 | 9995 | 10423 | 10642 | 10679 | 12313 | 11388 |
| 2023-12-22 | 9995 | 10423 | 10642 | 10679 | 12313 | 11388 |
| 2023-12-23 | 9995 | 10423 | 10642 | 10679 | 12313 | 11388 |
| 2023-12-24 | 9995 | 10423 | 10642 | 10679 | 12313 | 11388 |
| 2023-12-25 | 9995 | 10423 | 10642 | 10679 | 12313 | 11388 |
| 2023-12-26 | 9995 | 10423 | 10642 | 10679 | 12313 | 11388 |
| 2023-12-27 | 9995 | 10423 | 10642 | 10679 | 12313 | 11388 |
| 2023-12-28 | 9995 | 10423 | 10642 | 10679 | 12313 | 11388 |
| 2023-12-29 | 9995 | 10423 | 10642 | 10679 | 12313 | 11388 |
| 2023-12-30 | 9995 | 10423 | 10642 | 10679 | 12313 | 11388 |
| 2023-12-31 | 9995 | 10423 | 10642 | 10679 | 12313 | 11388 |

## IV. CONCLUSIONS AND RECOMMENDATIONS

Data on prices for lower quality rice I, lower quality rice II, medium quality rice I, medium quality rice II, super quality rice I, and super second quality rice originating from the National Strategic Food Price Information Center (PIHPS Nasional) for Banda Aceh is missing. value. This problem exists at several points in time, such as the price of rice on January 1, 2020, January 4, 2020, and January 5, 2020. Likewise, the price of rice in Banda Aceh is not recorded, including August 26, 2023 and August 27, 2023. So, this research solves this problem. with the last observation carried forward (LOCF) technique.

The complete data is then modeled using the auto-ARIMA technique using R programming. The recommended model for each rice quality forecast is ARIMA (0,0,5). The forecasting model for lower quality rice I has a constant of 9995.01, the first order moving average is 2.08, the second order moving average is 2.74, the third order moving average is 2.52, the fourth order moving average is 1.56, and the fifth order moving average is 0.58. Furthermore, the lower quality rice II forecasting model has a constant of 10422.90, the first order moving average is 1.73, the second order moving average is 2.11, the third order moving average is 1.97, the fourth order moving average is 1.37, and the fifth order moving average is 0.62. In addition, the medium quality rice I forecasting model has a constant of 10642.36, the first order moving average is 1.84, the second order moving average is 2.39, the third order moving average is 2.13, the fourth order moving average is 1.34, and the fifth order moving average is 0.52. For the medium quality rice II forecasting model, the constant is 10678.66, the first order moving average is 1.83, the second order moving average is 2.33, the third order moving average is 2.10, the fourth order moving average is 1.35, and the fifth order moving average is 0.52. After that, the super quality rice I forecasting model has a constant of 12313.12, the first order moving average is 1.63, the second order moving average is 1.95, the third order moving average is 1.79, the fourth order moving average is 1.29, and the fifth order moving average is 0.60. Finally, the super quality rice II forecasting model has a constant of 11387.50, the first order moving average is 2.08, the second order moving average is 2.79, the third order moving average is 2.54, the fourth order moving average is 1.59, and the fifth order moving average is the average is 0.60.

Based on the model's recommendations, the results of forecasting rice prices for all qualities show that there was a decline for some time (between September 1, 2023 and September 6, 2023) and then remained constant (between September 6, 2023 and December 31, 2023). The results of forecasting the price of lower quality rice I for the period September 1, 2023 and December 31, 2023 are between 9995 IDR and 11411 IDR. Furthermore, in the same period the forecast results for lower quality rice prices II were between 10423 IDR to 12005 IDR. Apart from that, the forecast results for the price of medium quality rice I for this time period are between 10642 IDR and 11995 IDR. Also, the forecast results for the price of medium quality rice II in that period are between 10679 IDR and 11945 IDR. Furthermore, the forecast results for the price of super quality rice I in that period are between 12313 IDR and 13640 IDR. Finally, the forecast results for the price of super quality rice I at that time were between 11388 IDR and 12761 IDR.

This research certainly has great potential for development. Other missing value techniques, both parametric and nonparametric, for example single imputation, maximum likelihood, multiple imputation, matrix factorization, and others can be tried for imputation of rice prices in Banda Aceh. Furthermore, the regions used can also be added so that the results of rice price forecasting can be compared between these regions. ARIMA models that have differencing also need to be tried and compared with the results in this study.